\documentclass[12pt]{article}

\usepackage{latexsym}
\usepackage{amssymb}
\usepackage{amsmath}
\usepackage{array}
\usepackage[english,italian,francais,german]{babel}
\newcommand{\bm}[1]{\mbox{\boldmath $#1$}}

\begin{document}

\centerline{\large \bf 
Absence of Zero Energy States}

\centerline{\large \bf in the Simplest d=3 (d=5?) Matrix
Models}

\vspace{0.4cm}
\centerline{Jens Hoppe}

\centerline{\small{Institute for Theoretical Physics, ETH H\"onggerberg,
    CH-8093 Z\"urich} } 
\vspace{0.4cm}

\centerline{Shing-Tung Yau}

\centerline{\small{Mathematics Department, Harvard University, Cambridge, MA
02138} } 

\vspace{0.4cm}

\centerline{Abstract}

\noindent The method introduced in [1] is simplified, and used to calculate the
asymptotic form of all SU(2) $\times SO(d=3$, resp. $5$) $invariant$ wave
functions satisfying $Q_{\hat{\beta}} \Psi = 0, \hat{\beta} = 1 \ldots 4$
resp. $8$, where $Q_{\hat{\beta}}$ are the supercharges of the $SU (2)$
matrix model related to supermembranes in $d+2=5$ (resp. $7$) space-time
dimensions. For $d=3$, there exist 2 asymptotic solutions, both of which are
constant (hence non-normalizable) in the flat directions, confirming previous
arguments that gauge-invariant zero energy states should not exist for $d<9$.
For $d=5$, however, out of 4 asymptotic singlet solutions (3 with orbital
angular momentum $l=0$, one having $l=1$) the one with $l=1$ does fall off fast
enough to be asymptotically normalizable, hence requiring further analysis to
be excluded as being extendable to a global solution.

\vspace{0.4cm}

As any of the bosonic degrees of freedom tends to infinity, each of the
hermitian supercharges
$Q_{\hat{\beta}}$, in the 4 possible matrix models $(d=2,3,5,7)$, may be
written as
$Q_{\hat{\beta}} = Q_{\hat{\beta}}^{(0} + Q_{\hat{\beta}}^{(1)} +
Q_{\hat{\beta}}^{(2)} + \cdots$
where $Q_{\hat{\beta}}^{(n+1)}$ is of order $r^{- \frac{3}{2}}$ smaller
than $Q_{\hat{\beta}}^{(n)}$, and $Q_{\hat{\beta}}^{(0)}$ commutes with
$r$ (the variable that
measures the distance from the origin in the space of configurations
having vanishing potential
energy). To leading and subleading order, $Q_{\hat{\beta}} \Psi = 0$, with
$\Psi = r^{- \kappa} ( \Psi_0 + 
\Psi_1 +\Psi_2 + \cdots)$ then gives
\begin{equation}
Q_{\hat{\beta}}^{(0)} \Psi_0 = 0 \label{e1} \end{equation}
\begin{equation}
Q_{\hat{\beta}}^{(0)} \Psi_1 + r^{\kappa} Q_{\hat{\beta}}^{(1)} r^{-
\kappa} \Psi_0 = 0 \quad .\label{e2}
\end{equation}
Asymptotic normalizability is governed by the
decay exponent $\kappa$, which follows (without having to calculate
$\Psi_1$) from projecting (2) onto any solution of (1), i.e. from

\begin{equation}
(\Psi_0^\prime , r^{\kappa} Q_{\hat{\beta}}^{(1)} r^{- \kappa} \Psi_0 ) =
0 \quad . \label{e3}
\end{equation}

Writing the bosonic variables in the form [1]
\begin{equation} 
q_{sA} = r e_A E_s + y_{sA} \quad , \label{e4}
\end{equation}
$A= 1,2,3, \quad s = 1, \ldots , d$ where $y_{sA} e_A = 0 = y_{sA} E_s , e_A
e_A = 1 = E_s E_s$, the leading and subleading (as $r \rightarrow \infty$)
terms in

\begin{equation}
Q_{\hat{\beta}} = \vec{\Theta}_{\hat{\alpha}} (-i \gamma^t_{\hat{\beta}
\hat{\alpha}}
\vec{\nabla}_t + \frac{1}{2} ( \vec{q}_s \times \vec{q}_t )
\gamma^{st}_{\hat{\beta} \hat{\alpha}}) \quad , \label{e5}
\end{equation}

when acting on SU (2) $\times SO (d)$ invariant wave functions $\Psi$, are
(cp. [1])
\begin{equation}
Q_{\hat{\beta}}^{(0)} = - i \Theta_{\hat{\alpha} A}
\gamma^{t}_{\hat{\beta} \hat{\alpha}} P_{AB} p_{st} \partial_{y_{sB}} + r
( \vec{e}\times \vec{y}_t ) E_s \gamma^{st}_{\hat{\beta} \hat{\alpha}}
\vec{\Theta}_{\hat{\alpha}} \label{e6} 
\end{equation}
\begin{equation}
Q_{\hat{\beta}}^{(1)} = - i \Theta_{\hat{\alpha} A}
\gamma^{t}_{\hat{\beta} \hat{\alpha}} (e_A E_t \partial_{r} + \frac{1}{r} E_t
M_{AB} e_B + \frac{1}{r} e_A M_{ts} E_s ) + \frac{1}{2} (\vec{y}_s \times
\vec{y}_t ) \gamma^{st}_{\hat{\beta} \hat{\alpha}}
\vec{\Theta}_{\hat{\alpha}}\quad , \label{e7} 
\end{equation}
with $P_{AB} : = ( \delta_{AB} - e_A e_B ), p_{st} 
: = ( \delta_{st} - E_s E_t ),$
\begin{equation*}
 \{
\Theta_{\hat{\alpha} A}, \Theta_{\hat{\beta} B}
\} = \delta_{\hat{\alpha} \hat{\beta}} \delta_{AB}
\end{equation*}
\begin{equation}
A,B = 1,2,3 \quad \hat{\alpha}, \hat{\beta} = 1, \ldots , s_d : = 4 \quad
\mathrm{(resp. 8);} 
\label{e8}
\end{equation}
$M_{AB} = \epsilon_{ABC} M_C$, resp. $M_{st}$, are the spin-parts of
the SU (2), resp.
$SO (d)$, generators
\begin{equation}
i J_A = \epsilon_{ABC} ( q_{sB} \nabla_{sC} + \frac{1}{2}
\Theta_{\hat{\alpha} B}
\Theta_{\hat{\alpha} C} ) \label{e9}
\end{equation}
\begin{equation}
i J_{st} = \vec{q}_s \vec{\nabla}_t - \vec{q}_t \vec{\nabla}_s +
\frac{1}{4}
\vec{\Theta}_{\hat{\alpha}} \gamma^{st}_{\hat{\alpha} \hat{\beta}}
\vec{\Theta}_{\hat{\beta}} \quad .\label{e10}
\end{equation}
The $s_d \times s_d$ dimensional $\gamma$-matrices are taken to be
\begin{eqnarray}
\gamma^d = \left( \begin{array}{cc} \bm{1} & 0 \\ 0 & -\bm{1} 
\end{array} \right) ,
\gamma^{d-1} = \left( \begin{array}{cc} 0 & \bm{1} \\ \bm{1} &
    0\end{array} \right) , \gamma^j = \left( \begin{array}{cc} 0 & i
    \Gamma^{j} \\ - i \Gamma^j & 0 \end{array} \right) \quad , \label{e11}
\end{eqnarray}
$\gamma^{st}:=\frac{1}{2}(\gamma^s\gamma^t-\gamma^t\gamma^s)$, with the
$\Gamma^j$ purely imaginary, antisymmetric, satisfying $\{ \Gamma^j
, \Gamma^k \} = 2 \delta^{jk} \bf{1}$.

For $d=5$ one could choose
\begin{equation}
\Gamma^1 = \sigma_1 \times \sigma_2 , \; \Gamma^2 = \sigma_2 \times \bf{1} ,
\; \Gamma^3 = \sigma_3 \times \sigma_2 \quad , \label{e12}
\end{equation}
and $\Gamma^1 = \sigma_2$ for $d=3$.

With the definition of the transverse annihilation operators,
$a_{\beta_{\nu}}$, given in [1], it is straightforward to verify that
\begin{equation}
\Psi_0 = e^{\frac{- r}{2} y^2} \mid F_{0}^{\perp} \rangle \mid
F_{0}^{\parallel} \rangle
\label{e13}
\end{equation}
satisfies $Q_{\hat{\beta}}^{(0)} \Psi_0 = 0$ if $\mid F_{0}^{\perp} \rangle =
{\prod a^{\dagger}_{\beta \nu`} \mid 0 \rangle_{x}}$, while  $\mid
F_{0}^{\parallel} \rangle$ can be any state
formed out of the
fermionic degrees of freedom $\Theta_{\hat{\alpha}}^{\parallel} : =
\vec{e}\cdot \vec{\Theta}_{\hat{\alpha}}$ and the bosonic variables $E_s$
(which, together with $r$ and $e_A$,
commute with $Q_{\hat{\beta}}^{(0)}$ ). The question is, what kind of
representations of $SO (d)$
the $2^{\frac{1}{2} s_d}$ dimensional ``parallel'' Fock space ${\cal H}$, with
creation operators
$\mu_{\alpha} : = \frac{1}{\sqrt 2} ( \Theta_{\alpha}^{\parallel} +
i \Theta^{\parallel}_{\alpha + {\frac{1}{2} s_d}} )$ contains.

The generators $M^{\parallel}_{st}$ of $SO(d)$ read
\begin{eqnarray}
M^{\parallel}_{d, d-1} &=& \frac{i}{2} ( \mu_{\alpha} \partial_{\mu_{\alpha}} -
\frac{1}{4} s_d ) \quad M^{\parallel}_{dj} = 
\frac{1}{4} \Gamma^{j}_{\alpha \beta} (
\mu_{\alpha} \mu_{\beta} - \partial_{\mu_{\alpha}} \partial_{\mu{_\beta}} )
\nonumber \\ [0.3cm]
 M^{\parallel}_{d-1,j} &=& \frac{-i}{4} \Gamma^{j}_{\alpha \beta} (
\mu_{\alpha} \mu_{\beta} + \partial_{\mu_{\alpha}} \partial_{\mu{_\beta}} )
\quad M^{\parallel}_{jk} = \frac{1}{2} \Gamma^{jk}_{\alpha \beta} \mu_{\alpha}
\partial_{\mu_{\beta} \quad .}\label{e14}  
\end{eqnarray}
Obviously, $\cal{H}$ splits into a direct sum of even and odd polynomials,
${\cal H}_{+} \bigoplus {\cal H}_{-}$, under
the action of (14).

For $d=3$, both basis elements of ${\cal H}_{-}$,
\begin{equation}
\mid F^{\parallel}_{0} \rangle^{(1)} = \mu_1 \mid 0 \rangle , \mid
F^{\parallel}_{0} \rangle^{(2)} = \mu_2 \mid 0 \rangle \label{e15}
\end{equation}
are annihilated by (14), while ${\cal H}_{+}$ is the representation space of a
spin $\frac{1}{2}$ representation of so(3) (over $\mathbf{C}$),
which cannot be matched (to give an overall singlet) by any representation
using the $E_s (s = 1,2,3)$. Hence there are exactly 2 singlet solutions
(asymptotically) for $d=3$.  Both of them give $\kappa=0$ (when using [1], one
may simply multiply equation (21) by 4, as for
$d=3\quad \Theta^{\parallel}_{\hat{\rho}} \; \Theta^{\parallel}_{\hat{\rho}} = 2$,
instead of 8; the contributions (42), (43) and (44) are then equal to $0, 1$,
and $-1$, resp., giving $\kappa = 0+1-1=0$).

Hence
\begin{equation}
\Psi^{(d=3)}_{0} = r^{-1} ( re^{- \frac{1}{2} ry^2} ) \mid F^{\perp}_{0}
\rangle \mid F^{\parallel}_{0} \rangle^{(1 or 2)} \label{e16} \end{equation}
which is not normalizable due to the radial measure $r^4 dr$ (the
$y=0$ manifold is
5-dimensional).

For $d=5$, the contributions analogous to (43) and (44) of [1] are $1 $
and $-2$, respectively
(having multiplied (21) by $2$, as $\Theta^{\parallel}_{\hat{\rho}}
\Theta^{\parallel}_{\hat{\rho}} = 4$); hence
\begin{equation}
\kappa_{d=5} = c_5 + 1 - 2 = c_5 -1 \label{e17} \end{equation}
where $c_5$ is the eigenvalue of
\begin{equation}
-\sum_{t=1}^{4} M^{\parallel}_{t5} M^{\parallel}_{t5} = -\frac{1}{2}
\sum_{t, s=1}^{5} (M^{\parallel}_{ts})^2 + \frac{1}{2} \sum_{\alpha , \beta
=1}^{4} (M^{\parallel}_{\alpha\beta} )^2, \end{equation}
when acting on $\mid F^{\parallel}_{0} \rangle_{E_s=\delta_{s5}}$. 
This time, ${\cal H}_+$
decomposes into a
5-dimensional representation of so(5), and 3 singlets, while ${\cal
H}_-$ splits into two 4-dimensional
representations of $so(5)$ $\cong sp (4)$. The 4 (overall singlet) states
\begin{equation}
\mid F^{\parallel}_{0} \rangle^{(j)} = \tilde{\Gamma}^{j}_{\alpha \beta}
\mu_{\alpha} \mu_{\beta} \mid 0 \rangle ,\; \mid F^{\parallel}_{0}
\rangle^{(4)} = E_s \mid s \rangle \quad ,
\label{e18}
\end{equation}
where
\begin{equation}
\tilde{\Gamma}^{1} = \sigma_2 \times \sigma_1 , \quad\tilde{\Gamma}^{2} =
\bm{1} \times \sigma_2 ,
\quad \tilde{\Gamma}^{3} = \sigma_2 \times \sigma_3 \label{e19} \end{equation}
and
\begin{equation}
\mid j \rangle = \frac{\sqrt 2}{4} \Gamma^{j}_{\alpha \beta} \mu_{\alpha}
\mu_{\beta} \mid 0 \rangle , \mid 4 \rangle = \frac{1}{\sqrt 2i}
( 1 + \mu_{1} \mu_{2} \mu_{3} \mu_{4} ) \mid 0 \rangle ,\; \mid 5 \rangle =
\frac{-1}{\sqrt 2} ( 1 - \mu_{1} \mu_{2} \mu_{3} \mu_{4} ) \mid 0 \rangle ,
\label{e20} 
\end{equation}
satisfying $M^{\parallel}_{st}\mid u\rangle = 
\delta_{tu}\mid s\rangle-\delta_{su}\mid
t\rangle$ ($\frac{1}{2}\epsilon_{\alpha\beta\gamma\delta}\Gamma^j_{\gamma
\delta}=-\Gamma^j_{\alpha\beta}$)
, then lead to 4 (asymptotic) singlet solutions,
\begin{equation}
\Psi^{( \alpha )}_{0} (d=5) = r^{-\kappa^{ (\alpha) } -2} ( r^{2}
e^{\frac{-1}{2}ry{^2}} ) \mid
F^{\perp}_{0} \rangle \mid F^{\parallel}_{0} \rangle^{ (\alpha) }
\label{e21}
\end{equation}
with
\begin{equation}
\kappa^{(j)} = -1 , \; \kappa^{(4)} = -1 +4 = +3 \label{e22} \end{equation}
i.e. effective fall-off $r^{-1} (l=0)$, resp. $r^{-5} (l=1)$. Given the radial
measure $r^{6} dr$, one finds that the $\Psi^{(j)}_{0}$ are not normalizable,
while $\Psi^{(4)}_{0}$ does fall off fast enough (hence further analysis is
needed to exclude the possibility that it may be extendable to a global
solution). Multiplying $r^{-5}$ by $r$ (the ratio of gauge variant to gauge
invariant radial measure, to the power of $\frac{1}{2}$), one gets, upon
multiplication by $E_s$, a function that is annihilated by the 5-dimensional
free Laplacian, resp. $\partial^{2}_{r} + \frac{4}{r} \partial_r - \frac{l
  (l+3)}{r^{2}}$, acting on a $l=1$ state (just as $( \partial^{2}_{r} +
\frac{4}{r} \partial_{r} ) (r^{-\kappa^{(j)} +1}) = 0$ for the three $l=0$
states). The asymptotic decay exponents $\kappa^{( \alpha )}$ are consistent
with [2], though not implied by their analysis of the asymptotics,
as the Fock space ${\cal H}$ of 'massless' fermions
$\Theta^{\parallel}_{\hat{\alpha}}$ (not treated in [2]) is needed, 
and -- for
fixed $l$ -- the choice which of the two possible eigenfunctions
of the free Laplacian (the decaying $r^{-l-d+2}$, or the non-decaying
$r^l$) is realized.

Finally, in order to check that $\Psi^{(4)}_{0}$ is consistent with
(3), one inserts $r^{-3} \Psi_{0} = \Psi^{(4)}_{0}, \Psi^{\prime}_{0} = e^{-
  \frac{r}{2} y^{2}} \mid F^{\perp}_{0} \rangle$, and multiplies by
$\gamma^{u}_{\hat{\rho} \hat{\beta}} E_u$, which gives the condition 
\begin{equation}
3 \Theta^{\parallel}_{\hat{\rho}} E_s \mid s \rangle = \Theta^{\parallel}_{\hat{\alpha}} ( \gamma^{u} \gamma^{t} )_{\hat{\rho} \hat{\alpha}}
E_u M^{\parallel}_{tv}
E_v E_s \mid s \rangle + \Theta^{\parallel}_{\hat{\rho}} E_s
\mid s \rangle -
\Theta^{\parallel}_{\hat{\rho}} E_s \mid s \rangle \frac{1}{\pi^{2}} \int_{-
\infty}^{+ \infty} e^{-ry^{2}} \frac{1}{2} ry^{2} d^{8}
(y \sqrt{r}) \label{e23}
\end{equation}
The term involving the integral contributes $-2$ (in [1], this would have
been $\frac{1}{2}$ (44)), so that (24) reduces to the identity
$\Theta^{\parallel}_{\hat{\alpha}} ( \gamma^{u} \gamma^{t} )_{\hat{\rho}
  \hat{\alpha}} E_u \mid t \rangle  = 4
\Theta^{\prime \prime}_{\hat{\rho}}  E_s \mid s \rangle$.

\vspace{0.4cm}

{\bf{Acknowledgement}}

One of us (J.H.) would like to thank O.~Augenstein and M.~Bordemann for
discussions, as well as H. Nicolai for correspondence.

\vspace{0.4cm} 
{\bf{References}}

[1] G. M. Graf, J. Hoppe; hep-th/9805020

[2] S. Sethi, M. Stern; hep-th/9705046

\end{document}